# A Variability-Aware Design Approach to the Data Analysis Modeling Process

Maria Cristina Vale Tavares
David R. Cheriton School of Computer Science
University of Waterloo
Waterloo, Canada
mvtavare@uwaterloo.ca

Paulo Alencar
David R. Cheriton School of Computer Science
University of Waterloo
Waterloo, Canada
palencar@uwaterloo.ca

Donald Cowan
David R. Cheriton School of Computer Science
University of Waterloo
Waterloo, Canada
dcowan@uwaterloo.ca

*Abstract*— The massive amount of current data has led to many different forms of data analysis processes that aim to explore this data to uncover valuable insights. Methodologies to guide the development of big data science projects, including CRISP-DM and SEMMA, have been widely used in industry and academia. The data analysis modeling phase, which involves decisions on the most appropriate models to adopt, is at the core of these projects. However, from a software engineering perspective, the design and automation of activities performed in this phase are challenging. In this paper, we propose an approach to the data analysis modeling process which involves (i) the assessment of the variability inherent in the CRISP-DM data analysis modeling phase and the provision of feature models that represent this variability; (ii) the definition of a framework structural design that captures the identified variability; and (iii) evaluation of the developed framework design in terms of the possibilities for process automation. The proposed approach advances the state of the art by offering a variability-aware design solution that can enhance system flexibility, potentially leading to novel software frameworks which can significantly improve the level of automation in data analysis modeling process.

## I. INTRODUCTION

The massive amount of current data has led to many different forms of data analysis processes that aim to explore this data to uncover valuable insights such as trends, anomalies and patterns. These processes support decision makers in their analysis of varied and changing data ranging from financial transactions to customer interactions and social network postings. These data analysis processes use a wide variety of methods, including machine learning and statistical data analysis, in several domains such as business, finance, health and operation of smart cities [1], [2]. In general, data analysis approaches have the potential to advance significantly the development of descriptive, diagnostic and predictive data analysis applications.

The process of data analysis involves phases such as data understanding, data preparation, modeling, evaluation and deployment. Specifically, the data modeling phase uses specific data analysis models to generate results in various situations. In general, this phase comprises tasks such as select modeling technique, generate test design, build model and assess model. Each of these tasks may numerous variations. For example, the selection of specific models to be applied depends on factors such as the application domain, business goal, suitability of the model type, data assumptions and available data types.

The high-level goal of this work is to make it easier to perform the data analysis modeling process, thus helping to support the development of big data science projects. To accomplish this goal there is a need to increase the extent to which this process can be automated. This study assumes that the analysis of the variability associated with the data modeling process can lead to capturing the variations in this process in a more comprehensive way and to producing a flexible software framework design. This design can help identify opportunities for automation that go beyond the automation support provided by existing tools. In this context, the work offers a variability-aware design approach to the data analysis modeling process.

*Problem*. From a software engineering perspective, the software automation of data analysis modeling processes faces numerous challenges. First, software users expect increased flexibility from the software in terms of the possible variations regarding different techniques, types of data, and parameter settings. The software is required to accommodate complex usage and deployment variations, which are difficult for non-experts. Second, variability in functionality or quality attributes increases the complexity in the design space of these systems and makes them harder to design and implement. There is a lack of a framework design that takes variability into account to support the design process. Third, the lack of a more comprehensive analysis of the variability design space makes it difficult to evaluate opportunities for automating the data analysis modeling process. This work addresses three main research questions:

(i) What is the variability related to the CRISP-DM data analysis modeling process and how can this variability be represented using feature models?
(ii) What structural framework design can capture the variability in the CRISP-DM data analysis modeling phase?
(iii) What are the opportunities for automation of the CRISP-DM data analysis modeling phase that go beyond the support provided by existing tools?

*Proposed Approach*. This study proposes a variability-aware design approach to the CRISP-DM data analysis modeling process, which involves (i) the assessment of the variability inherent in the CRISP-DM data analysis modeling phase; (ii) the definition of a framework design that captures the identified variability; and (iii) the evaluation of the proposed framework design in terms of the possibilities for process automation.

The assessment of the variability is based in the data analysis modeling phase of the CRISP-DM process model. The framework design represents the variability in the modeling process and highlights the complex dependencies among variation points and their variants. The evaluation of the framework design in terms of possibilities for automation can lead to enhancements in the level of automation supported by existing tools.

The proposed approach is illustrated in Fig. 1. First, a variability assessment is conducted using as sources the CRISP-DM documentation [3], [4] and articles that support the classification of the techniques that can be applied in the data analysis modeling process. Based on the variability assessment, feature diagrams are developed to represent the variability present in this process. Second, the feature diagrams are used to create a framework design for the CRISP-DM modeling phase. Third, the opportunities for automation of the CRISP-DM data analysis modeling phase that go beyond the support provided by existing tools are evaluated.

*Contributions.* In this study, we argue that the assessment of the variability of the data analysis modeling process can provide a framework design that can be used as the basis for evaluating opportunities for the automation of this process. The contributions of the study include:

(i) Variability assessment of the CRISP-DM data analysis modeling process, which provides variations related to its four generic tasks, namely Select Modeling Technique, Generate Test Design, Build Model and Assess Model, and the feature diagrams that represent these variations;
(ii) A framework design that is developed based on the four generic tasks of the variations identified in the variability assessment; and
(iii) An evaluation of the possibilities for automating the data analysis modeling process for each of the four generic tasks.

Overall, this work presents, to the best of our knowledge, the first approach based on variability assessment to design data analysis modeling processes, as defined in CRISP-DM model process. The approach advances the state of the art by offering a variability-aware design solution that can enhance system flexibility, potentially leading to novel software frameworks which can significantly improve the level of automation in data analysis modeling process.

## II. BACKGROUND AND RELATED WORK

### A. Data Analysis Processes

Data analysis is the process of applying approaches and computer systems to examine data in-depth, using tools, techniques, and methods, so that meaningful information can be identified to make decisions or solve problems [5]. The availability of digital data provided at a growing and rapid scale has led to new approaches of data analysis in order to turn data into useful knowledge, and improve decision making, create new business, and reduce costs [1], [6]. However, a manual data analysis of such huge volumes of data is not only impractical, but in some cases, almost an impossible task, which has raised the need for automating the data analysis modeling process.

Several data analysis processes have been proposed by industry and academia to describe the phases that data analysis experts. CRISP-DM (Cross-Industry Standard Process for Data Mining) and SEMMA (Sample, Explore, Modify, Model, and Assess) are the two most widely used in industry [3], [4], [7]. Specifically, CRISP-DM has modeling as one of its phases, which involves the decision on the most appropriate model for a given problem.

*CRISP-DM*. CRISP-DM is a standard data analysis process model and an associated methodology created in 2000 by a consortium of companies, (i.e., NCR, SPSS, and Daimler-Benz) [3]. The main goal for the proposition of this process model was to consolidate approaches and ideas related to the data analysis process through a standard, organized, and structured approach, meant to help beginners in data mining to understand the whole data analysis process and what to do to in each phase of this process.

As shown in Fig. 1, the data analysis process model proposed in CRISP-DM is a predefined sequence of six phases based on a cyclical approach [3], [4]. The arrows indicate the most common and meaningful sequences of phases that are typically applied in the data analysis process. However, moving back and forth between phases is expected in some cases [3], [4].

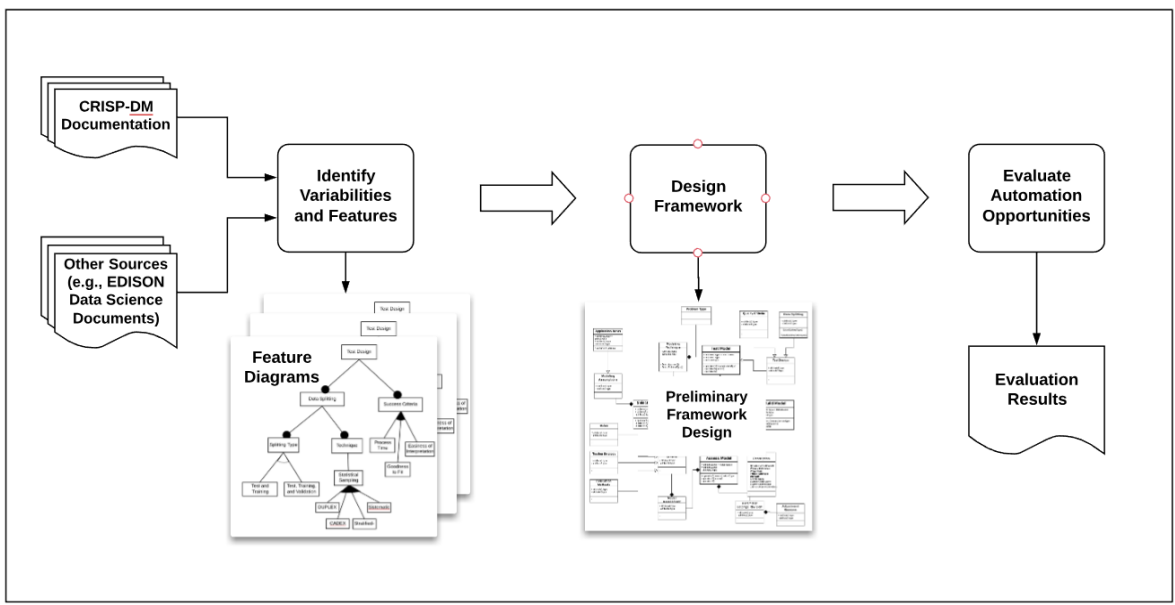

Fig. 1 A variability-aware design approach to the data analysis modeling process.

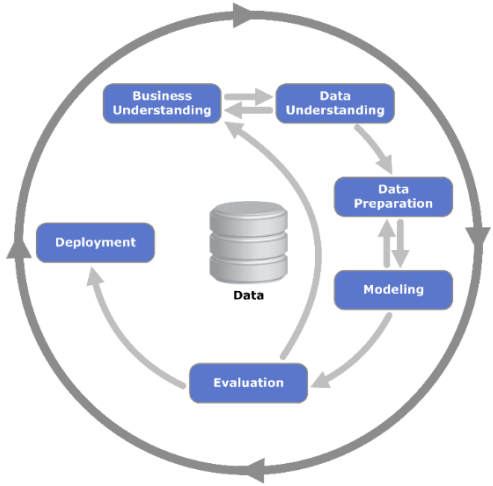

Fig. 2 CRISP-DM data analysis phases (from [9]).

*CRISP-DM Modeling Phase*. The data analysis modeling phase is an iterative process, in which several modeling techniques are applied to the same problem and their parameters are fine-tuned to optimum values until specific quality criteria is satisfied [4]. In the modeling phase, models are created through the application of modeling techniques and algorithms to extract interesting data patterns, correlations, and associations. This phase consists of four generic tasks:

**Select Modeling Technique** refers to the selection of a specific modeling technique, which should be performed separately for every technique applied, and comprises two outputs: the modeling techniques and the modeling assumptions. **Generate Test Design** refers to the generation of a procedure to test the model quality and validity needs, before building a model. In general, a dataset is divided into training and test sets, the model is built based on the training set, and its quality is estimated based on the test set. This task has only one output, the test design, which includes the description of the plan for training, testing and evaluating the models. **Build Model** refers to the creation of the models, running the modeling tool and using the prepared dataset as input. The outputs are the models produced by the selected tool, the parameter settings, and the model description. **Assess Model** refers to the assessment of the models based on the domain knowledge, the data analysis success criteria, the test design, and the decision on which models are both accurate enough and effective to be used in the next phases. The outputs of this task are: model assessment and the revised parameters settings.

*B. Variability Modeling and Analysis*

Variability is "*the ability of a software system or artifact to be efficiently extended, changed, customized or configured*" [8]. Variability applied to software engineering enables the adaptation of the structure, behavior, and development process of a software system, and helps to support the gradual evolution of a system software and enables innovation opportunities in system development [9]. Variability analysis can support the planning for future adjustments, adaptations and changes in a software system, making it reusable and customized in a variety of scenarios and domains. Over the system software life cycle, the variability analysis includes the identification of the software system parts that vary and the range of options for these variations, providing adaptability, flexibility, and reusability and facilitating the adaptation of a software system to changes, extensions, and customizations [9], [10], which implies that some decisions in the development process can be delayed; rather than de decided initially.

Studies on variability in software systems have been conducted and many variations of modeling approaches have been developed to implement variability techniques [10], [9], [11]. There are numerous types of variability models, including feature models [10], decision modeling [10], and models based on UML [12]. In general, variability can be assessed in many ways, one of which is by analyzing nouns and verbs in requirements system documents [13].

*Feature Models*. Commonality and variability of a product can be captured in an abstract way using entities called features [14]. According to Kang et al., a feature is "a prominent or distinctive and user-visible aspect, quality, or characteristic of a software system or systems" [14]. A feature model provides an abstract view of the variable and common requirements in a domain, enabling the definition of features, their properties, and their relations. Feature models help to support the development of common architecture and components, and constitute a key approach to plan for reusability. Feature models provides two types of relationship between features: one defines the relation between a parent feature and its child features and are represented by feature diagrams, and the other, called cross-tree relationship, defines the "requires" and "excludes" relationship in the model.

Feature models are graphically represented in a hierarchical diagrammatic notation, called a feature diagram, which expresses logical relationships among the features [14]. The connections between a parent feature and the group of child features are categorized as dependency relations that express possible composition rules, as presented in **Error! Reference source not found.** [15].

*UML Modeling*. For variable features modeled using UML, specifically UML class diagrams, the class diagrams and the variables can be derived manually using nouns and verbs [16]. In contrast, some techniques have been proposed for the automated generation of UML class diagrams from documents in natural language. Overall, these techniques still rely heavily on user input.

Table 1 Feature diagram notation.

| Notation | Description | Symbol |
|---|---|---|
| AND | All features must be selected | and |
| ALTERNATIVE | Only one feature can be selected | alternative |
| OR | One or more can be selected | or |
| MANDATORY | Features that are common to all instances | mandatory |
| OPTIONAL | Features that are optional | optional |

## C. Object-Oriented Frameworks

Over the years, software reuse has been studied and several techniques, tools, and applications have emerged [17]. The origin of the framework concept dates from mid-80s when it was defined as "a large structure that can be reused as a whole for the construction of a new system" [18] for a particular domain. A framework can also be seen as an object-oriented abstract design [19] that refers to "a generic architecture that provides an extensible template for applications within a domain" [15]. Software frameworks provide several benefits, including higher reuse and productivity and improved quality in the software system development process. The reuse of design patterns, components, and architectures is the foundation of the object-oriented reuse technique [17], enabling the customization and adaptation of applications within a specific domain.

## III. A VARIABILITY-AWARE DATA ANALYSIS MODELING PROCESS

The core results of this study as described in Fig. 1 are presented in this section: (i) Identify Variability and related Features; (ii) Design Framework; and (iii) Evaluate Automation Opportunities.

### A. Variability Assessment

The scope of the variability assessment is the modeling phase of the CRISP-DM model process, defined according to the CRISP-DM Reference Guide [3]. The variability assessment is conducted for each of the four generic tasks of the modeling process. The identification of variability and the features related to the CRISP-DM data analysis modeling phase are provided and these variables are represented as feature diagrams.

The assessment of variability starts with the analysis of the CRISP-DM documentation [3], [4] to understand the CRISP-DM data analysis process. Next, the modeling phase is comprehensively investigated (steps, generic and specialized tasks, activities, and outputs) to identify nouns, actions, verbs, and results of every task, and to evaluate potential variations for each one of these items. It was taken into consideration that an item might represent a variation point whenever it characterizes design options. For example, when we find "select method," because "method" is a noun, this can indicate that method can be a variation point that has as design options the several methods that can be adopted to the design. A variation point can lead to other variation points or ultimately to a final variant in the variation point hierarchy.

Part of the variability analysis, specifically the one related to the task "Select Modeling Technique", was mainly based on the EDISON Data Science documents [20], [21], a documentation developed in a research project called the EDISON Project. This project was created to establish a standardization in data related competences, such as professional skills, technological concepts, models, and knowledge areas. Our research takes into consideration the knowledge area classification for data analysis proposed in EDISON to support the assessment of variability related to approaches, methods, models, and algorithms for the task "Select Modeling Technique". Additional classifications provided in the literature contributed to enriching the variability analysis for the other tasks [22], [23].

The CRISP-DM Reference model [3] describes the modeling as the phase in which "various modeling techniques are selected and applied, and their parameters are calibrated to optimal values". The feature diagram for the whole modeling process represents the hierarchical structure of the refined properties for the modeling phase, whose concept feature is so called *Modeling*. The modeling process consists of four generic tasks, namely "Select Modeling Technique", "Generate Test Design", "Build Model" and "Assess Model" that are common characteristics of every modeling phase. Therefore, the four main features of the concept feature *Modeling* are defined as mandatory features: *Select Model Technique, Test Design, Build Model,* and *Assess Model*. Each one of these four features is described in the following paragraphs.

**Select Modeling Technique Feature**. The decomposition for part of this feature, presented in Fig. 3 described for the "Select Modeling Technique" task in [3], which involve: (a) the decision on which technique can be appropriately used, bearing in mind the tool selected; (b) the definition of any built-in assumptions made by the technique about the data (e.g., quality, format, distribution); and (c) the comparison and validation of the assumptions defined in the previous phases. These instances are required to occur, then both features *Modeling Technique* and *Modeling Assumptions* are related through a mandatory relationship with the feature *Modeling Technique*.

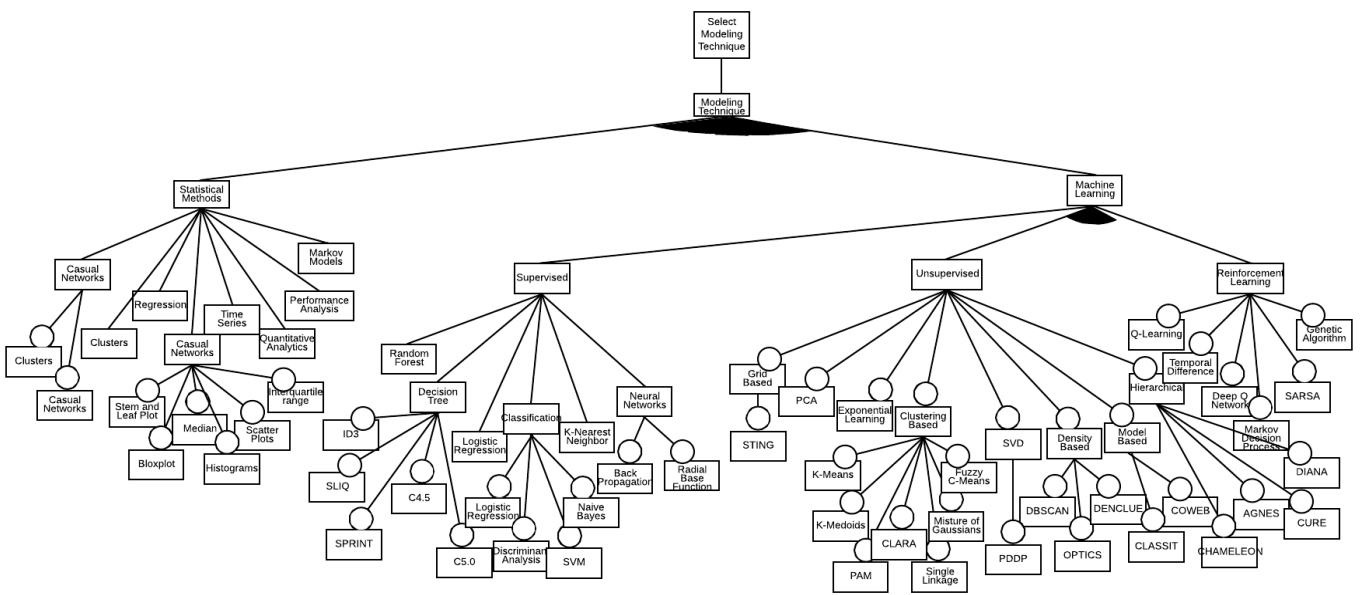

Fig. 3 Feature diagram for Select Modeling Technique feature.

The feature *Modeling Technique* is included in the diagram as an or-group feature. It expresses the process of selecting a modeling technique, which mainly involves deciding "on appropriate technique for exercise, bearing in mind the tool selected" [3].

In addition, according to [4], when deciding on which model(s) to use, some issues should be considered, such as; (a) data splitting technique; (b) data availability to produce reliable results for a given model; (c) quality level of current data; (d) data type appropriateness for a particular model; and (e) data conversion needs. Thus, choosing a technique to generate a model takes into consideration elements that may be combined into several multiple categories such as data analysis modeling approach, modeling technique types, model types, and algorithm types.

The EDISON project [20] provides data analysis (so-called data science) subject domain classifications based on existing standard, commonly accepted approach categorizations, and other publications from industrial and research communities. In this context, it proposes created data analysis knowledge areas and related sub-classifications. Based on that structure, the first level of or-group feature corresponds to the data analysis modeling approach namely *Statistical Methods, Machine Learning, Data Mining, Predictive Analytics, Computational Modeling Techniques,* and *Domain Analytics Methods*. The second level is also based on the EDISON project [20] and refers to modeling technique types. For example, the feature *Machine Learning* is a parent feature whose child features depict the group of modeling technique types which include types such as Supervised, Unsupervised, and Reinforced Learning. The third level of child features represents the model types. For example, for the feature *Supervised*, some of features include *Decision Tree, Naive Bayes, Ordinary Least Square Regression, Logistic Regression, Neural Networks, SVM, Ensemble Methods*, and others. The last child feature refers to the algorithm type, and some examples represented in the feature diagram are the *Apriori Algorithm, ID3, C5.0,* and *SVM*.

The mandatory feature *Modeling Assumptions*, not shown in Fig. 3, consists of a group of alternative features which specify assumptions about data according to the modeling technique selected [3]. The activities for this step involve [3]: (a) defining any built-in assumptions made by the technique about the data (e.g., quality, format, and distribution); (b) comparing these assumptions with those defined previous phases; and (c) ensuring that the assumptions hold and go back to the previous phase, if necessary. Determining an appropriate model would also need to consider the availability of data types for mining, the data mining goals, and the specific modeling requirements [4]. These elements are optional and might occur depending on the technique selected. Based on this, the feature diagram depicts or-group features, namely: *Data-related Assumptions, Application areas,* and *Model Data Type*.

*Test Design*. To generate the feature sub-diagram for "Test Design", the description of the "Generate Test Design" task was considered. This task describes the procedure of testing the model results in terms of quality and validity. This procedure is applied to as many models as needed and depends on the number of models selected to be used and deployed. The output of this step is a comprehensive test design that is "the intended plan for training, testing, and evaluating the models" [3]. This test design describes "the criteria for goodness of a model" and defines "the data on which these criteria will be tested" [4]. As stated in [3], the activities for this step involve checking the existing and appropriate test designs for each data mining goal, preparing data required for the test, and deciding on necessary steps, such as number of iterations or number of folds.

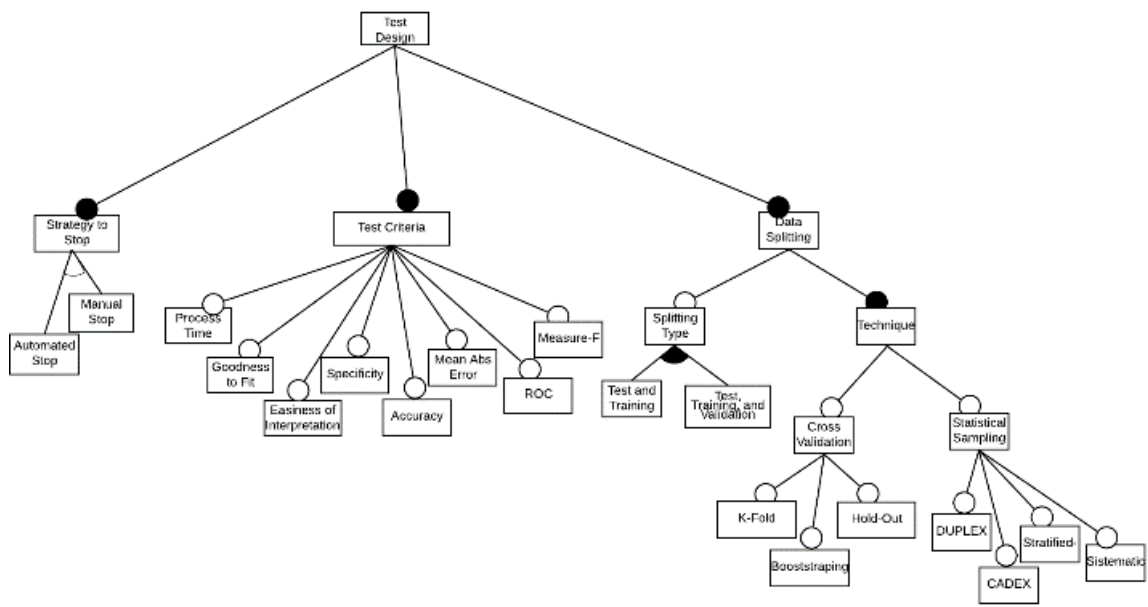

Fig. 4 Feature diagram for Test Design feature.

Based on the previous description of the "Generate Test Design" task, the feature *Test Design* consists of five mandatory features: *Quality Criteria, Data Splitting Technique, Data Splitting, Strategy to Stop,* and *Success Criteria*. Fig. 4 shows the diagram related to the feature *Test Design*.

The feature *Quality Criteria* was defined as a mandatory feature because "it is necessary to define a procedure to test the model's quality", as stated in [3]. The quality of a model is measured based in characteristics that contribute to its quality. A wide range of quality measures have been defined in the literature [49], [69], [70], such as sensitivity, accuracy, specificity, the ROC curve, and Mean Absolute Error. These examples of metrics form an or-group feature, indicating that one or more can be selected as a metric for model quality measurement: *Sensitivity, Accuracy, Specificity, the ROC Curve,* and *Mean Absolute Error*.

Another feature is *Success Criteria*, which consists of features of the ways a model goodness can be measured. The success criteria may vary according to the modeling technique. For example, "for unsupervised models, such as Kohonen cluster nets, measurements may include criteria such as ease of interpretation, deployment, or required processing time" [4]. These features also form an or-group feature.

*Strategy to Stop* is another mandatory feature, which is based in the fact that "modeling is an iterative process, it is important to know when to stop adjusting parameters and try another method or model" [4]. For this feature, two alternatives features are derived: *Automated Stop* and *Manual Stop*, of which only one can be selected.

To introduce the Data Splitting feature, it was considered that in the "Generate Test Design" step, a procedure is used to "separate the dataset into train and test sets, build the model on the train set, and estimate its quality on the separate test set" [3]. The need to separate datasets into subsets for different indicates that Data Splitting is a mandatory feature. The primary component of the plan for testing the model involves determining how to partition the available dataset into training, test, and validation datasets [3], [4]. Thus, the data split options form another alternative-group feature, namely Test and Training, or Test, Validation, and Training.

There is a number of techniques that can be applied to the data splitting process. The *Cross Validation* and *Statistical Sampling* form an or-group feature that describes common approaches to data splitting methods. The feature *Cross Validation* contains child features that represent commonly used types of cross-validation strategies, such as Hold-Out, K-Fold, and Bootstrapping [24]. The feature *Statistical Sampling* represents some of most widely used statistical sampling approaches, including: DUPLEX, CADEX, Stratified Sampling, Simple Random Sampling, Convenience Sampling, and Systematic Sampling [24].

***Build Model Feature***. The mandatory feature Build Model expresses the variability identified in the task "Build Model" as described in the CRISP-DM documentation. Fig. 5 shows the diagram related to the feature Build Model. This task refers to the creation of one or more models by a modeling tool to support the data mining decisions by comparing the results and analyzing the notes that were made during the process [3], [4]. It is important to track the progress of a variety of models, keeping "notes of the settings and data used for each model" [4]. By the end of the model-building process, three outputs will be produced: "Parameter Settings", "Model Description", and "Models" [3]. These elements are represented in the diagram as two mandatory features *Parameter Settings* and *Model Generation*, and an optional feature named *Model Description*.

The output "Parameter Settings" is a list consisting of information about "the initial parameters" and "the reasons for choosing those values" [3], [4]. Therefore, two mandatory features are represented in the diagram to express these two activities involved in the parameter settings [3], which are named *Settings* and *Rationale for Choosing*. According to [3], "there are often a large number of parameters that can be adjusted" and, therefore, considering that parameters can be initially set and later adjusted, the feature *Settings* consists of two features related to the moment the setting of parameters might occur. These features are expressed as *Initial* and *Adjustments*. In its turn, the initial parameters can be set automatically or manually, which are represented in the diagram as features *Manual* and *Automated*. To set parameters automatically, it can be pre-set with default settings parameters or parameters related to the technique applied. Therefore, two exclusive features have relationship with feature *Automated Default Setting* and *Technique Parameter*.

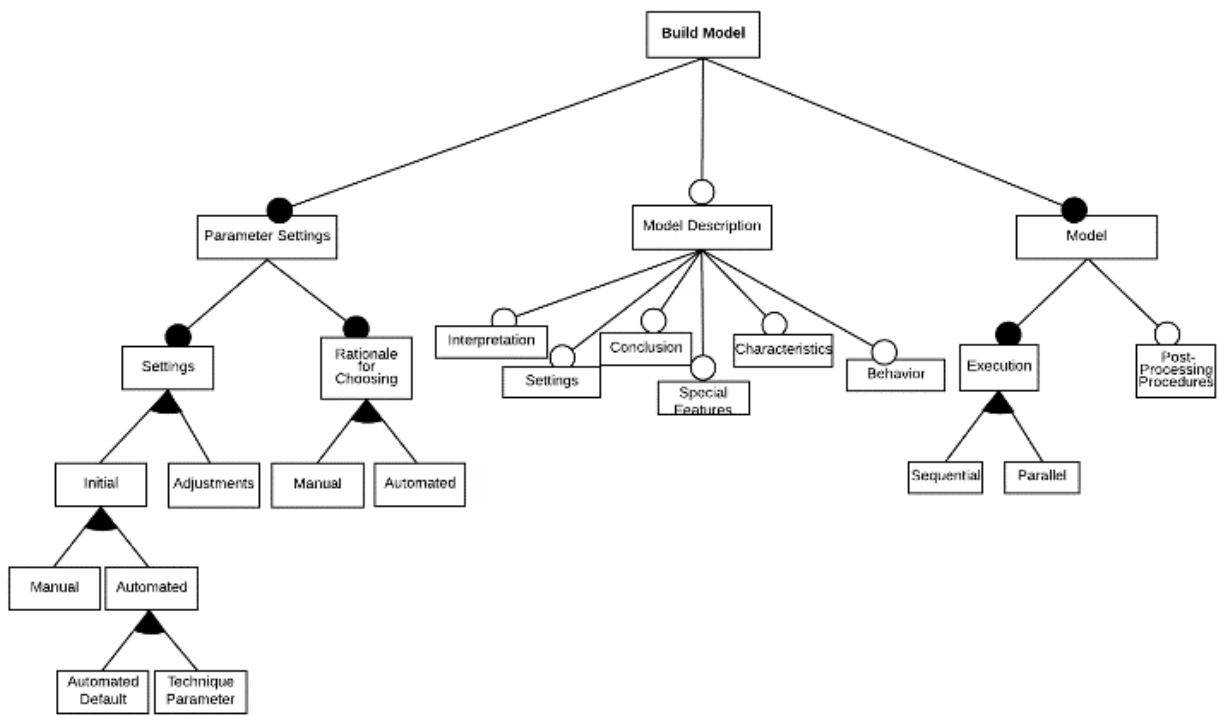

Fig. 5 Feature diagram for Build Model feature.

The activity of taking notes of parameter choices and reasons for the choice [3] can be performed manually or automatically. This activity is expressed as the feature *Reasons*, consisting of two exclusive features: *Manual* and *Automated*.

"Model" is another output of the "Build Model" step and refers to the creation of one or more models from the modeling tool. The activities in this task are running the selected technique on the input dataset to produce model and post-processing data mining results (i.e. edit rules, display trees) [3]. The feature *Model* represents the output "Model" and consists of two features named *Execution* and *Post-processing Procedures* to express the activities performed for building a model. The feature *Execution* refers to creation of the models and contains two features *Sequential* and *Parallel*, which express the way the modeling creation may occur. The feature *Post-processing Procedures* represents the post-processing results of the model generation.

The last output is the "Model Description" which is expressed in the diagram as an optional feature, also named Model Description. According to [3], when examining the results of a model, it is recommended to record information about the modeling experience and its meanings. Activities performed in this task are: (a) describing of characteristics of the current model that may be useful for the future; (b) recording parameter settings used to produce the model; (c) providing a detailed description of the model and any special features; (d) listing the rules produced, plus any assessment of per-rule or overall model accuracy and coverage or list of any technical information about the model (such as neural network topology) and any behavioral descriptions produced by the modeling process (such as accuracy or sensitivity); (e) describing the model's behavior and interpretation; and (f) providing conclusions regarding patterns in the data (if any); sometimes the model reveals important facts about the data without a separate assessment process (e.g., that the output or conclusion is duplicated in one of the inputs) [3]. In the feature diagram, these activities and properties were expressed as an optional-or feature group consisting of Interpretation, Parameter Settings, Conclusion, Special Features, Characteristics, and Behaviors.

*Assess Model*. The last task of the modeling phase is "Assess Model". The model assessment refers to the analysis of the effectiveness and accuracy of the results to define the final models that will be deployed. The evaluation is based on the criteria generated in the test plan to ensure the model meets success criteria [3], [4]. The activities involved in the model assessment include [3], [4]: (a) evaluating results regarding evaluation criteria; (b) testing result according to a test strategy; (c) comparing the results and interpretation of evaluation; (d) ranking results; (e) selecting best models, (f) interpreting results in business terms; (g) getting comments on models by domain or data experts; (h) checking plausibility of model and effect on data mining goal; (i) checking the model against given knowledge base to see if the discovered information is novel and useful; (j) checking the reliability of the result; (k) analyzing potential for deployment of each result; (l) assessing the rules (in terms of logic, feasibility, quantity, etc) and the results; and (m) getting insights into why a certain modeling technique and certain parameter settings lead to good/bad results. Fig. 6 shows the diagram that represents the feature *Assess Model*. According to [3], this task generates two outputs: the "Model Assessment" and the "Revised Parameter Settings", which are represented in the feature diagram as features *Model Assessment* and *Revised Parameter Settings*.

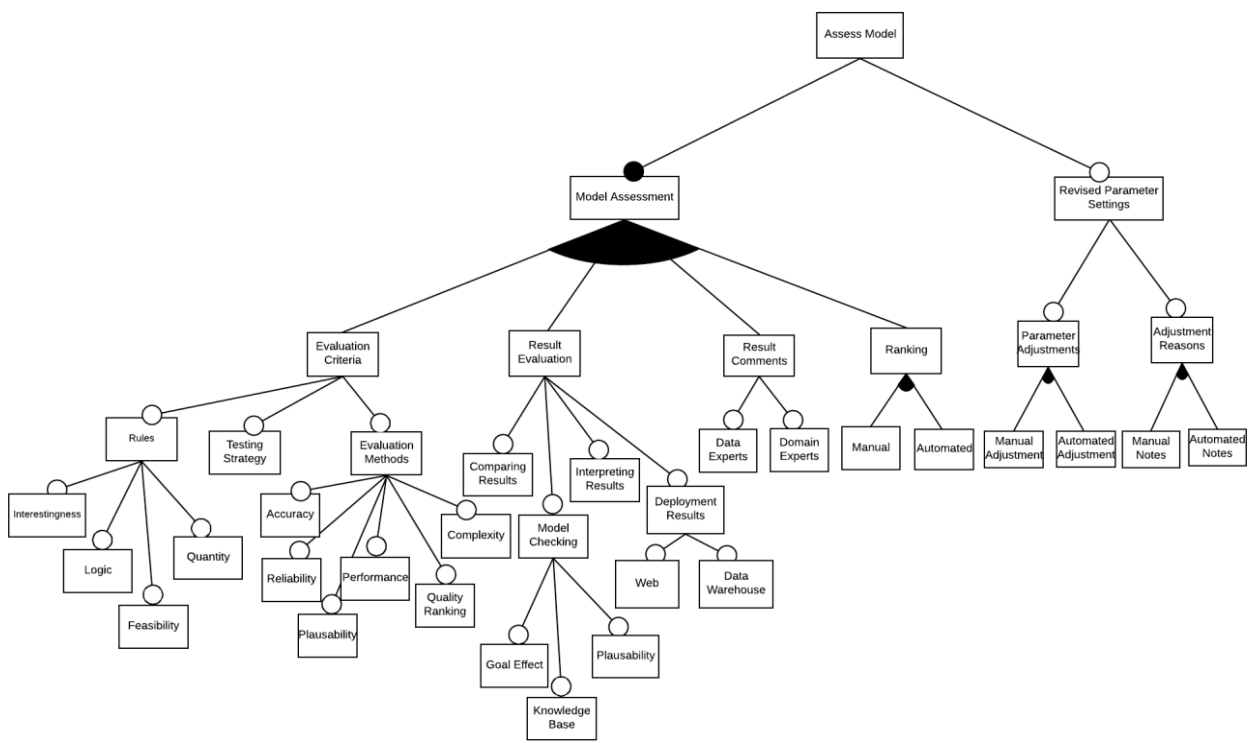

Fig. 6 Feature diagram for Assess Model feature.

The "Model Assessment" is a summary of the results, the qualities of the generated model, and the quality rank (in relation to each other) [3]. Taking into account the activity descriptions, the main aspects of the model assessment are expressed in the diagram as three features called *Evaluation Criteria, Results Evaluation, Result Comments,* and *Ranking*.

The second output, "Revised Parameter Settings", is generated from an iterative process of adjusting, revising and tuning parameter settings [3]. The main activities refer to revise parameters initially set and record the reasons for adjusting them. These activities are represented in the diagram as two mandatory features, which are called *Parameter Adjustment* and *Adjustment Reasons*. Both activities can be performed manually or in an automated way. Therefore, the ways for performing the adjustment of the parameters are expressed as alternative features named *Manual Adjustment* and *Automated Adjustment*. In its turn, the feature *Adjustment Reasons* can also be executed in two different ways and is represented in the diagram as two alternative feature named *Manual Notes* and *Automated Notes*.

B. *Framework Design Model*

In this section, based on variability analysis of the CRISP-DM modeling phase, a framework design to support the CRISP-DM data analysis modeling phase is defined using a UML class diagram. This framework design is showing in Fig. 7.

This framework design represents the variability associated with the CRISP-DM data analysis modeling phase for each of its tasks, subtasks and outputs. These variability, which were derived informally using nouns and verbs [16], are modeled using UML, specifically UML class diagrams [16]. Each variation point is defined as a class in the UML class diagram and the variation points or variants associated with a specific variation point are defined using the sub-class relationship.

The class CRISP-DM Modeling Controller refers to the CRISP-DM Modeling phase and consists of four general sub-systems that represent the four generic tasks of the modeling phase. These components are named *Modeling Technique Selector, Test Design Generator, Model Builder,* and *Model Evaluator*.

The *Modeling Technique Selector* represents the first step of the Modeling phase. It provides a class named *Model Technique* that represents the feature *Modeling Technique*. This class is composed of the classes *Technique* and *Assumption* which are the representation of the variation points identified previously. The class *Technique* expresses the variation points and variants related to a technique such as the approach, technique type, model type, and algorithm. To exemplify an instance of Technique, Machine learning can be classified as an approach, Supervised Machine Learning as a type of technique, Decision Tree as a type of a model, and C5.0, a type of an algorithm. The class *Assumption* represents the possible variability of assumptions resulted from the technique selected, such as data format, data type, splitting strategy, and others.

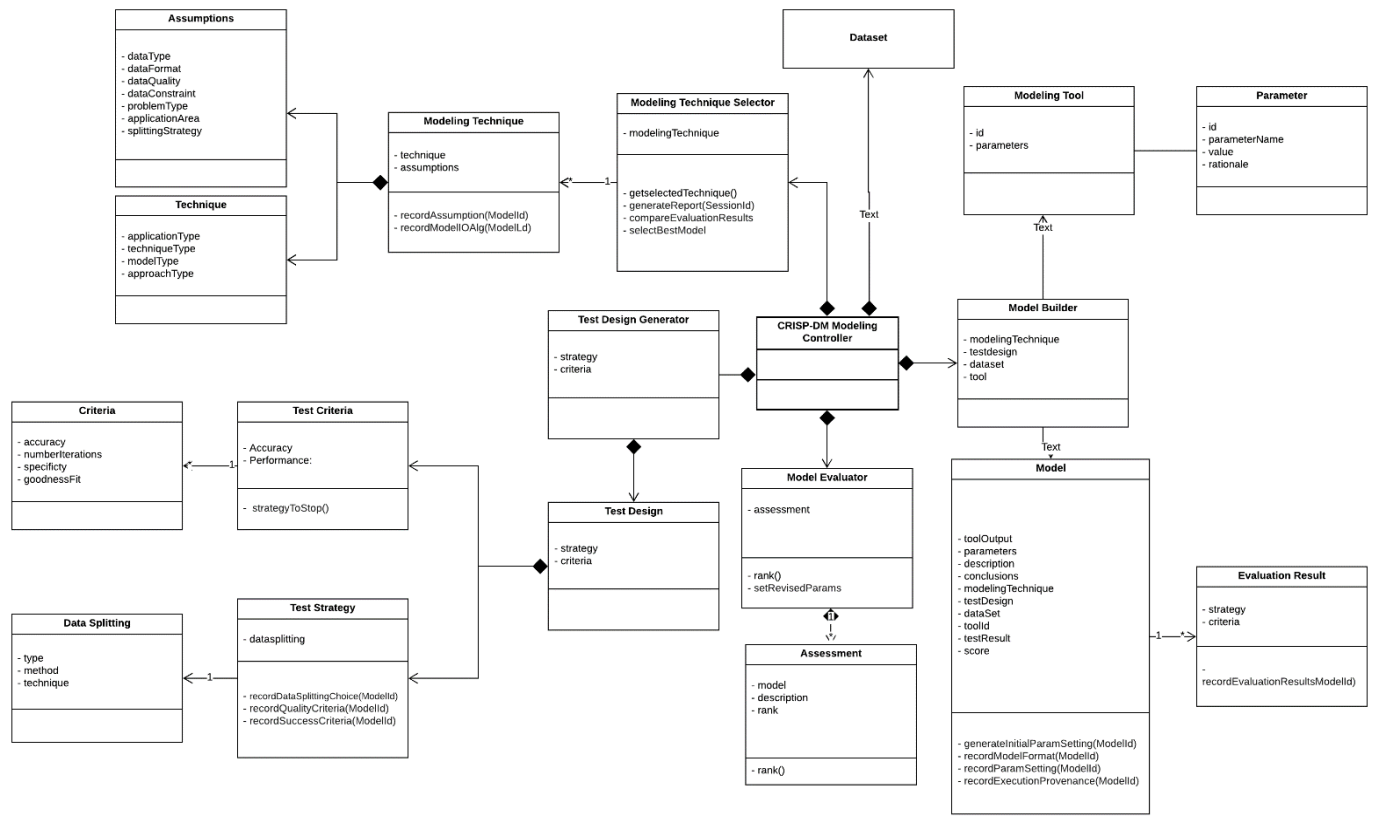

Fig. 7 Framework design for CRISP-DM modeling phase.

The *Test Design Generator* represents the second step and provides a class *Test Design,* which refers to variations identified in the feature diagram for the feature Test Design. The class *Test Design* is composed of the classes *Test Criteria* and *Test Strategy* which are the representation of the variation points identified for the Test Design step. The class *Test Criteria* expresses the variation points and variants related to the procedure to test the model's quality and validity. The class *Test Strategy* represents the variability of the data splitting process. It is composed of a class named Data Splitting which specifies the type, method, and techniques that may be applied in the splitting decisions.

The *Model Builder* represents the third step and the variability identified in the variability assessment. It provides the classes *Model* and *Model Tool*. The class *Model* represents variability related to the model, such as the parameters, score, conclusions, and description. *Model Tool* represents the variability of the tool used for building the model, such as parameters. Not all tools use the same set of parameters to build a model.

*Model Evaluator* refers to the fourth step and the associated variability. The assessment step considers strategy and criteria defined previously to provide the model rank. Classes *Assessment* and *Evaluation Result* represent the variability identified in the process of assessing a model. *Assessment* refers to the description of the result of the evaluation and the model rank, while *Evaluation Result* represents the criteria and the strategy for the model evaluation.

*C. Evaluation of Automation Opportunities*

The framework design described previously open up several opportunities for automation in the data analysis modeling process. This section discusses these automation opportunities in each of the four CRISP-DM data analysis modeling tasks by comparing the proposed framework structural design with the design solution provided in the SPSS Modeler - CRISP-DM [4].

We have chosen to compare the framework design with the SPSS Modeler for CRISP-DM to assess automation opportunities for several reasons. First, SPSS Modeler is the only tool we have found that aims at automating the CRISP-DM data analysis process. Second, this tool has been used in many applications throughout the years, including applications in domains such as finance, marketing and smart cities. Third, although we did not have access to SPSS Modeler CRISP-DM requirements document, we could access a significant amount of information provided in the online user manual. Overall, based on the variability that can be captured by our framework structural design, there are many opportunities for automating the CRISP-DM data analysis modeling process that go beyond the support provided by the existing SPSS Modeler for CRISP-DM. These opportunities for automation, which can significantly increase the level of automation of the CRISP-DM modeling process, include:

(i) The ability to support a more comprehensive set of modeling techniques and algorithms.
(ii) Mechanisms to record the assumptions required by each technique and algorithms in a more detailed way.
(iii) The ability to record the types of the data used by the models as input and output and the association with the algorithms that used them.
(iv) The ability to record and support data splitting choices, a more comprehensive set of quality criteria and success criteria;
(v) Mechanisms to support the generation of initial parameter settings.
(vi) The ability to record the rationale for parameter value choices.
(vii) Additional mechanisms for supporting interoperability in terms of the methods and input and output results.
(viii) The ability to record the resulting model characteristics, parameter settings, data quality issues, and provenance related to the data analysis method execution (e.g., what, who, when).
(ix) Mechanism to rank the models.
(x) Mechanisms for generating revised parameter settings.

The following sections provide the comparison between SPSS Modeler CRISP-DM [4] and the framework design in terms of variability support for each one of the four generic tasks in the modeling phase.

*Select Modeling Technique*. The Select Modeling Technique phase is divided into two subtasks: Modeling techniques and Modeling assumptions. For Modeling techniques, the comparison of the variability support provided by the proposed framework design with the automation features provided in the SPSS Modeler CRISP-DM indicates that there are several opportunities for automation of this task. The SPSS Modeler for CRISP-DM tool does not cover all the automation possibilities that can be supported by our framework design, including:

(i) The ability to support a more comprehensive set of modeling techniques, categorized by approach, technique type, model type, and algorithms.
(ii) Mechanisms to record the assumptions required by each technique and algorithms in a more detailed way. The set of a variety of types of data-related assumptions (e.g., specific modeling requirements, strategies for data splitting, data manipulations needed to meet the model requirements) can be useful for non-expertise users.
(iii) The ability to record the types of the data used by the models as input and output and the association with the algorithms that used them.
(iv) The incremental record of the data used by the models.
(v) The ability to associate fields of application with techniques and assumptions.

*Generate Test Design*. The Generate Test Design phase is structured in only one subtask: Test design. The findings indicates that SPSS Modeler CRISP-DM tool does not support all possible variability that can be supported by our design approach, such as:

(i) The ability to encode data splitting techniques and record data splitting choices.
(ii) The ability to encode and record a more comprehensive set of quality and success criteria.
(iii) The ability to encode strategies to stop.
(iv) The possibility of evaluating and comparing results of every operational model.

*Build Model*. The Build Model phase is divided into two subtasks: Parameter settings; Models; and Model description.

The findings related to the comparison indicates that SPSS Modeler CRISP-DM tool does not support variability such as:

(i) The generation of initial parameter settings.
(ii) The provision of additional mechanisms for supporting interoperability in terms of the methods and input and output results that can be used by other tools.
(iii) The ability to record the resulting model characteristics, parameter settings, data quality issues.
(iv) The ability to record provenance related to the data analysis method execution (e.g., what, who, when).

*Assess Model*. The Assess Model phase is divided into two subtasks: Model assessment and Revised parameter settings. The findings related to the comparison of the design solution for this subtask with the SPSS CRISP-DM support tool indicates the tool does not support variability such as:

(i) The ability to record detailed evaluation criteria, in terms of the rules, testing strategy, and evaluation methods.
(ii) The provision of several methods to assess the model according to recorded description type (e.g., provenance, parameter settings).
(iii) The ability to record task result summary and key steps of the evaluation process.
(iv) The ability to record the comments from data and domain experts.
(v) The ability to rank model quality.
(vi) The provision of additional techniques to support the selection of the best model.
(vii) The ability to support decision on parameter revision and the ideal number of iterations.

## IV. CONCLUSIONS AND FUTURE WORK

This study proposes a variability-aware design approach to the data analysis modeling process. The proposed approach helps to advance the state of the art by providing potential design enhancements to existing solutions and indicating novel ways to automate the data analysis modeling process. These results are beneficial both to designers and practitioners who are involved in the design and implementation of processes to help support big data science projects.

Future work related to this study may involve: (i) the refinement of the proposed framework design; (ii) the development of specific case studies in particular domains; (iii) the extension of the approach to other data analysis processes and other phases; (iv) the use of (knowledge) databases to capture data relevant to the process; and (v) the provision of automated learning capabilities that can provide user guidance through recommendations.